\documentclass[manuscript, nonacm=true]{acmart}
\usepackage{import}
\usepackage{algorithm} 
\usepackage{algpseudocode} 
\usepackage{amsthm}
\algdef{SE}[SUBALG]{Indent}{EndIndent}{}{\algorithmicend\ }%
\algtext*{Indent}
\algtext*{EndIndent}
\newtheorem{lemma}{Lemma}[section]
 \newtheorem{innercustomlemma}{Lemma}
\newenvironment{customlemma}[1]
  {\renewcommand\theinnercustomlemma{#1}\innercustomlemma}
  {\endinnercustomlemma}
\theoremstyle{definition}
\newtheorem{definition} {Definition}

\newif\ifdraft
\draftfalse
\newcommand{\arxiv}[1]{\ifdraft{\color{red}[{\bf Added Feedback}: #1]}\fi}

\usepackage{tlatex}

\AtBeginDocument{%
  \providecommand\BibTeX{{%
    \normalfont B\kern-0.5em{\scshape i\kern-0.25em b}\kern-0.8em\TeX}}}

\begin{document}

\title{On the Significance of Consecutive Ballots in Paxos}

\author{Eli Goldweber}
\email{edgoldwe@umich.edu}
\affiliation{%
  \institution{University of Michigan}
  \city{Ann Arbor}
  \state{Michigan}
}
\author{Nuda Zhang}
\email{nudzhang@umich.edu}
\affiliation{%
  \institution{University of Michigan}
  \city{Ann Arbor}
  \state{Michigan}
}
\author{Manos Kapritsos}
\email{manosk@umich.edu}
\affiliation{%
  \institution{University of Michigan}
  \city{Ann Arbor}
  \state{Michigan}
}




\begin{abstract}
  In this paper we examine the Paxos protocol and demonstrate how the discrete numbering of ballots can be leveraged to weaken the conditions for learning. Specifically, we define the notion of consecutive ballots and use this to define Consecutive Quorums. Consecutive Quorums weakens the learning criterion such that a learner does not need matching $accept$ messages sent in the \textit{same ballot} from a majority of acceptors to learn a value. We prove that this modification preserves the original safety and liveness guarantees of Paxos. We define \textit{Consecutive Paxos} which encapsulates the properties of discrete consecutive ballots. To establish the correctness of these results, we, in addition to a paper proof, formally verify the correctness of a State Machine Replication Library built on top of an optimized version of Multi-Paxos modified to reflect $Consecutive \; Paxos$.
\end{abstract}




\maketitle


\section{Introduction}\label{Intro}
More than 20 years after its inception, the Paxos algorithm~\cite{PartTimeParliment, Paxos} remains a fundamental building block for distributed consensus and State Machine Replication (SMR) in an asynchronous setting. The importance of Paxos is made evident by the numerous variants of the algorithm --- e.g.~\cite{DiskPaxos,RingPaxos,FlexPaxos,FastPaxos,GeneralisedConsensus,EPaxos} --- and its use in real-world deployments \cite{libpaxos,Spanner,Megastore}. Despite two decades of research on this topic, however, we have yet to understand all the subtleties of the Paxos algorithm.

The correctness of Paxos can be expressed concisely as ``no two different values can be learned''. This simple property, in turn, relies crucially on the way that values are learned: 
\begin{itemize}
\item A value can only be learned if a majority of acceptors accepts it {\em in the same ballot}. 
\end{itemize}

The above criterion for learning is considered fundamental to the correctness of Paxos. And yet, this paper demonstrates that it is, in fact, {\em stronger} than it needs to be. Before we formally state and prove our claim, we will illustrate how this criterion is stronger than necessary by observing some specific examples of Paxos in action.

\subsection{Consecutive Quorums}

Let us consider a number of Paxos snapshots, as seen from the perspective of a learner. Each snapshot shows the state of the five acceptors in a Paxos ensemble with $f=2$. For each acceptor, we show the accepted value and the ballot in which this value was accepted. Since the learner only needs to receive an $accept$ message from a majority of acceptors, we use ``?'' to denote the state of acceptors for which the learner has not received an $accept$ message. For each snapshot, we consider the following question: {\em is it safe to learn value $x$?}


\begin{table}[ht]
\parbox{.45\linewidth}{
\centering
\begin{tabular}{|c|c|c|}
\hline
{\textbf{Acceptor ID}} &
{\textbf{Value}}& 
{\textbf{Ballot}}  \\
\hline 
A & ? & ? \\
\hline 
B & ? & ? \\
\hline 
C & $x$ & 10 \\
\hline 
D & $x$ & 9 \\
\hline 
E & $x$ & 7 \\
\hline 
\end{tabular}
\caption{Information collected at the learner. According to Paxos, value $x$ cannot be learned yet. This is correct, since another value can still be learned.}
\label{tab:non-consecutive}

}
\hfill
\parbox{\linewidth}{
\centering
\begin{tabular}{|c|c|c||c|c||c|c||c|c|}
\hline
{\textbf{Proposer Number}} & 
\multicolumn{2}{|c||}{7} & 
\multicolumn{2}{|c||}{8} &
\multicolumn{2}{|c||}{9} &
\multicolumn{2}{|c|}{10} \\
\hline
{\textbf{Acceptor ID}} &
{\textbf{Value}}& 
{\textbf{Ballot}} &
{\textbf{Value}}& 
{\textbf{Ballot}} &
{\textbf{Value}}& 
{\textbf{Ballot}} &
{\textbf{Value}}& 
{\textbf{Ballot}} \\
\hline 
A & - & - & - & -& - & -& - & -\\
\hline 
B & - & - & $y$ & 8 & $y$ & 8 & $y$ & 8 \\
\hline 
C & - & - & - & -& - & -& $x$ & 10 \\
\hline 
D & - & -& - & -& $x$ & 9 & $x$ & 9 \\
\hline 
E & $x$ & 7& $x$ & 7 & $x$ & 7 & $x$ & 7 \\
\hline 
\end{tabular}
\caption{Illustration of a possible execution leading up to the state observed by the learner in Table~\ref{tab:non-consecutive}. This shows the state of each acceptor after proposers numbered $7 \dots 10$ have proposed a value. Proposer \#11 can safely propose value $y$ after receiving a quorum of promise messages \{$A$(-,-), $B$($y$,8), $E$($x$,7)\}, thus eventually causing value $y$ to be learned.}
\label{tab:stateSequence}
}

\end{table}

Table~\ref{tab:non-consecutive} illustrates an example where the learner knows that a majority of acceptors have accepted the same value, but not in the same ballot. One might be tempted to conclude that, since a majority has already accepted $x$, it is henceforth impossible for any other value to be learned---and thus it is safe to learn $x$. But this is not the case. Consider the following execution (illustrated in Table~\ref{tab:stateSequence}). Initially, no acceptor has accepted a value. The proposer with ballot number 7 (henceforth, proposer \#7) proposes $x$, which is accepted by $E$. Then proposer \#8, who happens not to hear from $E$ during phase one, is elected and proposes value $y$, which is accepted by acceptor $B$. At this point, proposer \#9 performs phase one of Paxos and receives promise message from acceptors $C$ (-,-), $D$(-,-), and $E$($x$,7), and thus proposes value $x$, which is accepted by $D$. Similarly, proposer \#10 receives promise messages from $C$(-,-), $D$($x$,9), and $E$($x$,7), and proposes value $x$, which is accepted by $C$. At this point, it is {\em not} safe to learn $x$, since it is still possible for a subsequent proposer, say proposer \#11, to receive promise messages from $A$(-,-), $B$($y$,8), and $E$($x$,7), and to thus propose $y$. 

The above reasoning seems to lend credence to the criterion used in Paxos to determine whether a value can be learned: ``a majority of acceptors must accept a value {\em in the same ballot}''. A majority of acceptors that accept the same value in different ballots is not enough to merit learning that value. But what if the ballots are all {\em consecutive}?

Table~\ref{tab:consecutive} shows just such an example, where a majority of acceptors have accepted the same value, $x$, in consecutive ballots. In this paper we claim that in this case, $x$ {\em can} be learned. The intuition behind this claim comes from observing the example of Table~\ref{tab:non-consecutive}. The reason why $x$ cannot yet be learned in that example is that there exists a ballot (\#8 in this case) with a different value $y$, whose ballot number supersedes one of the ballots in the majority that has accepted $x$. This makes it possible that a future proposer picks this value $y$ as its proposed value. When the majority consists only of consecutive ballots, however, no such ``interleaved'' ballot exists, which makes a {\em consecutive quorum} just as strong as if all ballots had the same ballot number. 

\begin{table}[t]
\parbox{.45\linewidth}{
\centering
\begin{tabular}{|c|c|c|}
\hline
{\textbf{Acceptor ID}} &
{\textbf{Value}}& 
{\textbf{Ballot}}  \\
\hline 
A & ? & ? \\
\hline 
B & ? & ? \\
\hline 
C & $x$ & 10 \\
\hline 
D & $x$ & 9 \\
\hline 
E & $x$ & 9 \\
\hline 
\end{tabular}
\caption{Information collected at the learner. According to Paxos, value $x$ cannot be learned yet. And yet, {\em no other value} can be learned.}
\label{tab:consecutive}
}

\end{table}

We therefore propose the following weakening of Paxos --- and all its corresponding variants. Given a Paxos ensemble with $2f+1$ acceptors, a learner can learn a value $x$ as soon as it receives $f+1$ $accept$ messages from distinct acceptors if: (a) all such messages denote acceptance of value $x$, and (b) the set of the ballot numbers of these messages consists of consecutive numbers.

\subsection{Formal verification}

Section~\ref{consecQSection} includes a proof showing that our proposed weakening does not affect the safety of Paxos. To validate our confidence in the correctness of our reasoning, we use formal verification techniques to produce a mechanically-checked proof showing that this weaker version of Paxos still maintains the safety property of the original Paxos algorithm.

\subsection{Contributions}
In summary, this paper we make the following contributions:
\begin{enumerate}
    \item We show that the criterion for {\em learning} values in the original Paxos algorithm is too strong. We propose a weaker criterion that we call \textbf{Consecutive Quorums}.
    \item We prove that this weakened criterion maintains the same correctness guarantees provided by the original Paxos algorithm.
    \item We incorporate the Consecutive Quorums idea into a weaker version of Paxos that we call \textbf{Consecutive Paxos}. We use formal verification techniques to build a mechanically-checked proof showing that Consecutive Paxos maintains the safety property of the original Paxos algorithm.
\end{enumerate}
The rest of the paper is organized as follows. Section~\ref{related_work} discusses related work. Section \ref{model} formalizes our model of Paxos and introduces our notation. Section~\ref{consecQSection} introduces Consecutive Quorums and proves that this weakening maintains the safety guarantees of Paxos and does not adversely affect liveness. Section~\ref{formalVerification} discusses our methodology for formally verifying the correctness of Consecutive Paxos. Section~\ref{DiscussionSection} discusses additional ways in which consecutive ballots can affect the design of Paxos and Section~\ref{conclusion} concludes the paper. A TLA+ model for the combined proof of all techniques using consecutive ballots can be found in Appendix~\ref{apendix:consecP+Q}.

\section{Related Work}\label{related_work}
Distributed consensus, and especially Paxos, has been extensively studied. Many existing works aim to optimize various aspects of Paxos. Some variations improve throughput, decrease latency, increase fault tolerance, or apply consensus to specific environments \cite{FastPaxos,RingPaxos,DiskPaxos,GeneralizedPaxos,Mencius}. There has also been considerable work explaining Paxos and showing how it can be applied in practical settings \cite{PaxosLive,Paxos,DeconstructingPaxos,ComplexPaxos,ABCDs}.


Most of these variations use the structure of Paxos as a building block, but some have focused on the foundational principles. Flexible Paxos \cite{FlexPaxos} shows that overlapping majority quorums between phases is not a necessary condition in Paxos. Safety is preserved as long as quorums used for learning values intersect with quorums used for proposing values, regardless of size.
In \cite{GeneralisedConsensus,ConsensusRevised}, Howard explores the reasoning for how Paxos solves consensus. This work presents ways in which Classic Paxos is stricter than needed. In addition to relaxing the quorum intersection requirement, Howard explains how receiving a $promise$ message gives the proposer implicit information about previous rounds.

In addition, WPaxos \cite{WPaxos} makes use of flexible quorums in a geo-replicated setting. Generalized Paxos \cite{GeneralizedPaxos}, $M^2$Paxos \cite{M2Paxos}, Egalitarian  Paxos \cite{EPaxos}, and CAESAR \cite{CAESAR}, increase flexibility and performance by leveraging commutative commands or allow for any replica to act as leader. These Paxos-inspired variations re-examine how to exploit different attributes in the protocol, quorum sizes, re-ordering independent commands, and split leadership, to help generalize consensus. None of these works, however, leverage the structure of consecutive ballots, and thus they could all benefit from the ideas presented in this paper.  


Paxos solves consensus in an asynchronous setting but is not the only solution. View-stamped Replication was proposed earlier than Paxos \cite{viewstamped}. Raft \cite{Raft} was designed to solve the same problem, but with an emphasis on the practical setting. Paxos remains the most studied consensus protocol in the academic setting, as evident by the plethora of papers that build on top of it. To gain a better understanding of core differences between Raft and Paxos, Wang et al. \cite{RaftToPaxos} studied the root differences between the two. The resulting work shows a refinement mapping from a modified version of Raft to Paxos. This work demonstrates how to use refinement to map several Paxos variants to the world of Raft. 


The general structure of Paxos shares many similarities with protocols designed to reach consensus in the presence of Byzantine agents. PBFT \cite{PBFT} and XFT \cite{XFT} along with the many other variations reflect the general structure of Paxos, values are decided by replicas exchanging messages in discrete ballots over a series of phases. Values are proposed and learned in much the same manner as Paxos. More recent work has applied the quorum insights from Flexible Paxos to BFT \cite{FlexByzantine}. Hotstuff \cite{Hotstuff} relies on changing views for each proposal. The advantage of pipelining consecutive views allows Chained Hotstuff to reduce the overall number of messages needed to reach consensus. The insight of discrete consecutive ballots is general to a wide range Paxos based protocols even those that tolerate Byzantine failures. 

\section{Model and Overview}\label{model}
\subsection{Paxos}

We consider the model of Classic Paxos \cite{Paxos}. A system consisting of
\begin{itemize}
    \item a collection of \textit{proposer} agents that propose values,
    \item a collection of \textit{acceptor} agents that accept values, and
    \item a collection of \textit{learner} agents that learn values.
\end{itemize}

Agents communicate by passing messages over an asynchronous network. In typical implementations, agents are mapped to processes, where each process consists of one proposer, acceptor and learner agent.

In the Paxos protocol, proposers associate each proposed value $v$ with a ballot number $n$, and acceptors accept proposals that are ballot-value pairs. We denote such a proposal as \proposal{n}{v}. It is important for different proposals to have different ballots, which can be achieved by each proposer using ballots from its own disjoint pool. Where relevant, subscripts denote the source of each message. We now give a description of the Classic Paxos protocol.

\textbf{Phase 1.}
\begin{enumerate}[(a)]
    \item A proposer $\pi$ selects a ballot number $n$ and sends a \prepare{n}{\pi} request to the acceptors.
    \item Upon receiving \prepare{n}{\pi}, an acceptor $\alpha$ promises not to respond to any more proposals numbered less than $n$. If $n$ is the greatest ballot $\alpha$ has promised, it then responds to $\pi$ with \promise{n}{\proposal{n'}{v'}}{\alpha}, where \proposal{n'}{v'} is the highest-numbered proposal $\alpha$ has previously accepted. If $\alpha$ has not yet accepted any proposals, it then sends a promise message containing a special null symbol \promise{n}{\bot}{\alpha}.
\end{enumerate}

\textbf{Phase 2.}
\begin{enumerate}[(a)]
    \item If a proposer $\pi$ receives responses from a majority of acceptors in response to its prepare request numbered $n$, the proposer sends \propose{n}{v}{\pi}, where $v$ is the value of the highest numbered proposal in the responses, or any value if all responses contained $\bot$.
    \item Upon receiving \propose{n}{v}{\pi}, an acceptor $\alpha$ promises not to respond to any more proposals numbered less than $n$. If $n$ is the greatest ballot $\alpha$ has promised, it then accepts the proposal $(n,v)$ and sends \accept{n}{v}{\alpha} to the learners.
    \item A learner learns that a value is chosen if it receives \textit{accept} messages from a majority of acceptors with the same value and ballot.
\end{enumerate}

In Classic Paxos, a value $v$ is \textit{chosen} if there exists a majority quorum of acceptors that have accepted the same proposal $(n, v)$, that is, the same value $v$ in the same ballot $n$. Once a value is chosen, two things occur. First, any pending proposals from a lower ballot will be ignored by at least a majority of acceptors, any previous proposal can no longer achieve majority acceptance. Second, any higher numbered proposal must have a Phase 1 quorum that intersects with the quorum of acceptors that accepted the chosen proposal. By the protocol specification, this can only result in the proposer proposing the same $v$ in Phase 2. Along with the requirement that any ballot be associated with a single value, guarantees that once a value is chosen, no different value can ever be chosen or learned.

In particular, Paxos ensures that:
\begin{itemize}
    \item Only a value that is proposed can be chosen,
    \item Only a single value can be chosen, and
    \item Only a value that is chosen can be learned.
\end{itemize}

Given $2f+1$ acceptors, Paxos is safe given the absence of malicious failures, and live given no more than $f$ acceptors fail by crashing, during sufficiently long periods of synchrony \cite{FLP}.

\subsection{Our Approach}
%

In principle, the set of all ballots used in Paxos can be any unbounded, ordered set, including innumerable ones. However, it is ubiquitous in both theory \cite{Paxos, FlexPaxos} and practice \cite{libpaxos,Spanner,Megastore,Chubby} to use as ballots the set of natural numbers, or any likewise discrete set. The additional structure that the natural numbers provide can be used to weaken the criterion for learning values in Paxos:
\begin{itemize}
    \item \textbf{Consecutive Quorums} It is not necessary that a quorum of acceptors accept a value \textit{in the same ballot} for the value to be learned. A value can be chosen---and eventually learned---as long as a majority of acceptors accept that value in an unbroken span of \textit{consecutive} ballots.
\end{itemize}


\begin{definition}
\label{consecPairDef}[Consecutive Pair] 
We define two ballots $m \leq n$ in $\nat$ as consecutive iff there does not exist any ballot $m'\in\nat$ such that $m<m'<n$. 

$$Consecutive(m,n) \equiv m \leq n \; \land \; \nexists m' \in \nat.\; m<m'<n $$
\end{definition}

\begin{definition}
\label{consecSetDef}[Consecutive Set] 
A set $S \subseteq \nat$ is said to be consecutive iff for any element $n$ in $S$, it is either the smallest element, or there is an element $m$ smaller than $n$ in $S$ such that $m$ and $n$ are consecutive. 

$$ConsecutiveSet(S) \equiv \forall n \in S .\; (n = min(S)) \lor (\exists  m \in S .\; m <  n \land Consecutive(m,n))$$
\end{definition}

The $Safety$ of Paxos is the property \textit{"no two different values can be learned"}. A value is \textit{learned} after a learner receives a quorum of \textit{accept} messages indicating that a value has been chosen. To specify these properties, we let each correct Paxos agent $\theta$ maintain two sets as its persistent state, $sent_\theta$ and $msgs_\theta$. The set $sent_\theta$ contains exactly the messages sent by $\theta$. The set $msgs_\theta$ contains exactly the messages received by $\theta$. The safety of Paxos inherently relies on intersecting quorums of acceptors, denoted as $Q$, that contain at least $f+1$ of the $2f+1$ total acceptors. A valid majority quorum formed in ballot $i$ is expressed as $Q_i$.

In Paxos, the Safety property is:

\begin{theorem}[Safety Property]
\label{safetyTheorem}
No two different values can be learned
\[ \forall i,j \in \nat .\, [Learned(i,v)
                 \land Learned(j,w)] \implies v = w \]
\end{theorem}

where $Learned$ is defined as follows:

\begin{definition}
\label{learnedDef}[Learned] 
Value $v$ is \textit{learned} in ballot $i$ iff a learner, $l$, receives $accept$ messages for value $v$ in ballot $i$ from a majority of acceptors. 

$$Learned(i,v) \equiv \exists Q_i .\, \forall \alpha \in Q_i .\, \accept{i}{v}{\alpha}  \in msgs_l$$
\end{definition}

To maintain the Safety Property, Paxos ensures \textit{"no two different values can be chosen."} A learner does not learn a value until a value is chosen. If this invariant holds, then it must be the case that\textit{"no two different values can be learned"}. Using this definition, the following is a formal description of the Chosen Invariant: 

\begin{theorem}[Chosen Invariant]
\label{ChosenInvarant}
No two different values can be chosen
\[ \forall i,j \in \nat .\, [Chosen(i,v)
                 \land Chosen(j,w)] \implies v = w \]
\end{theorem}

In Classic Paxos, the following is the criterion that designates if a value is \textit{chosen}:

\begin{definition}
\label{chosenDef}[Chosen] 
Value $v$ is \textit{chosen} in ballot $i$ iff a majority of acceptors send matching $accept$ messages. 

$$Chosen(i,v) \equiv \exists Q_i .\, \forall \alpha \in Q_i .\, \accept{i}{v}{\alpha}  \in sent_\alpha$$
\end{definition}

By proving the Chosen Invariant \ref{ChosenInvarant}, the Safety Property \ref{safetyTheorem} follows directly.



\section{Consecutive Quorums} \label{consecQSection}
Consecutive ballots can be leveraged to weaken the criterion of how values are chosen and learned in Paxos. In this section we prove that Consecutive Quorums --- which consist of a majority of acceptors that have accepted a value across a consecutive set of ballots --- provide the same safety guarantee as majority quorums in Classic Paxos. By ensuring that no two different values can be chosen, at most one unique value can be learned.



\begin{definition}
\label{consecQ}[Consecutive Quorum (CQ)] 
A quorum of acceptors $Q = \{ \alpha_1, \dots, \alpha_n\}$ is considered a \textit{Consecutive Quorum} supporting the value $v$ iff the set of ballots from the sent \textit{accept} messages form a \textit{Consecutive Set}
\begin{align*}
&CQ(Q,v) \equiv\; |Q| \geq f+1 \\
& \qquad \qquad \; \; \land \exists \accept{x_1}{v}{\alpha_1} \in sent_{\alpha_1} \land \dots \land \exists \accept{x_n}{v}{\alpha_n} \in sent_{\alpha_n} \\
& \qquad \qquad \; \; \land ConsecutiveSet(\{x_1, \dots, x_n\})
\end{align*}

\end{definition}

Using this definition, we can weaken the criteria for a value to be considered \textit{Chosen} in ballot $i$ to: 

\begin{definition}
\label{consecQChosen}[CQ Chosen] 
\textit{A value is chosen in ballot} $i$ once there is a majority quorum of acceptors that have sent $accept$ messages with matching values and the ballots form a consecutive set. Additionally, at least one $accept$ message must be sent in ballot $i$.

\[ Chosen(i,v) \equiv \exists Q_i .\; CQ(Q_i,v) \land \exists \alpha \in Q_i.\; \accept{i}{v}{\alpha} \in sent_{\alpha}  \]

\end{definition}
Note that \textit{CQ Chosen} is strictly weaker than the Classic Paxos definition of \textit{Chosen}. A classic majority quorum of matching $accept$ messages is just a specific case of a Consecutive Quorum where all the ballots are of the same number.

The safety of Classic Paxos relies on the intersection of quorums. As long as a single quorum of acceptors sends matching $accept$ messages to a learner, the chosen value is set in stone. Any future proposer, before proposing any value, will first obtain a quorum of $promise$ messages. At least one acceptor will participate in both quorums. This ensures that only the chosen value could be proposed in a future ballot. A learner learns that value $v$ is \textit{CQ Chosen} in ballot $i$ once receiving $accept$ messages from a majority of acceptors, with matching values and the ballots form a consecutive set. Additionally, at least one $accept$ message must be sent in ballot $i$.




Learning with Consecutive Quorums does not affect the safety of Paxos. Consecutive Quorums are still majority quorums that will intersect with all other quorums. However, the ballot associated with the acceptor in the intersection might be different than in Classic Paxos without Consecutive Quorums. A valid Consecutive Quorum contains acceptors that have accepted the same value but from potentially different ballots. If the intersecting acceptor did not have the highest numbered proposal from the Consecutive Quorum, the definition of consecutive ballots ensures that there cannot exist any ballots between the reported ballot and the highest ballot in the Consecutive Quorum that could have a different value. As a result, \textit{CQ Chosen} guarantees that once a value is chosen, no different value could also be chosen.

As an example, consider a set of 5 acceptors. For Classic Paxos to consider value $v$ as chosen in ballot $i$, at least 3 acceptors must send $accept$ messages for $v$ in ballot $i$. In the case where the learner observes $accept$ messages from distinct acceptors for value $v$ in ballots $i-2$, $i-1$, and $i$, Classic Paxos cannot learn a value. However, This constitutes a valid Consecutive Quorum, and $v$ would be considered learned. No other value could possibly be learned at this point. 

\subsection{Safety Proof For Consecutive Quorums}
To prove the safety of Consecutive Quorums, we must show that the Safety Property \ref{safetyTheorem} cannot be violated when using the updated definition for choosing a value, \textit{CQ Chosen}. To prove that Safety Property \ref{safetyTheorem} holds, we prove the \textit{Chosen Invariant} \ref{ChosenInvarant} with the following three Lemmas over all ballots $i,j \in \nat$.

\begin{lemma}
\label{lemma1Q}[Equal CQ]
If $v$ is chosen in ballot $i$ then no other value $w \neq v$ could be chosen in any ballot $j = i$
\end{lemma}

\begin{lemma}
\label{lemma2Q}[Less Than CQ]
If $v$ is chosen in ballot $i$ then no other value $w \neq v$ could be chosen in any ballot  $j < i$
\end{lemma}

\begin{lemma}
\label{lemma3Q}[Greater Than CQ]
If $v$ is chosen in ballot $i$ then no other value $w \neq v$ could be chosen in any ballot  $j > i$
\end{lemma}

By showing that Lemmas \ref{lemma1Q}, \ref{lemma2Q}, and \ref{lemma3Q} hold, it is straight forward to prove the safety of Paxos with Consecutive Quorums. The \textit{Chosen Invariant} holds for all ballots $i,j$; the above three Lemmas cover all cases, when $i = j, i < j,$ and $i > j$. Therefore Lemma \ref{lemma1Q} $\land$ Lemma \ref{lemma2Q} $\land$ Lemma \ref{lemma3Q} implies \textit{Chosen Invariant}. In turn, \textit{Chosen Invariant} implies Safety Theorem. 

To help prove the above three Lemmas, we make use of the following definitions that are derived directly from the design of Paxos:
\theoremstyle{definition}
\begin{definition}
\label{LearnImpliesAccept}[Chosen Implies Accept]
If value $v$ is Chosen in ballot $i$, then $v$ must have been Accepted by at least one acceptor in ballot $i$.
\end{definition}

\begin{definition}
\label{AcceptImpliesPropose}[Accept Implies Propose]
If value $v$ is Accepted in ballot $i$ (by some acceptor), then $v$ must have been Proposed in ballot $i$.
\end{definition}

We can combine Definitions  \ref{LearnImpliesAccept} and \ref{AcceptImpliesPropose}, using the transitivity of implication, to define a third equivalent definition.

\begin{definition}

\label{LearnImpliesPropose}[Chosen Implies Propose]
If value $v$ is Chosen in ballot $i$, then $v$ must have been Proposed in ballot $i$.
\end{definition} 

To prove Lemma \ref{lemma1Q}, we can first prove Lemma \ref{lemma1Qa}. A proof of Lemma \ref{lemma1Qa} along with the Definition \ref{LearnImpliesAccept} (\textit{Chosen Implies Accept}), proves, using the transitivity of implication, Lemma \ref{lemma1Q}. 
\begin{customlemma}{4.1'}
\label{lemma1Qa}
If an acceptor has accepted $v$ in ballot $i$ then no other value $w \neq v$ could be chosen in any ballot $j = i$
\end{customlemma}

\begin{proof}
Assume at least one acceptor has accepted value $v$ in ballot $i$. There must be at least one acceptor that has accepted $v$ in ballot $j$. From Definition \ref{AcceptImpliesPropose} (\textit{Accept Implies Propose}), we know that $v$ was proposed in ballot $j$. To finish this proof, we can rely on the construction of Paxos itself. Only a single unique value can be proposed per ballot. $v$ was already proposed in ballot $i$, so no other value could be proposed in ballot $j$. A value cannot be learned if it was not proposed. Proposing $v$ in ballot $j = i$ cannot lead to the formation a valid Consecutive Quorum where $w \neq v$ could possibly be learned in ballot $j = i$.
\end{proof}

To prove the \textit{Chosen Invariant} for all ballots $i,j \in \nat$, we consider two cases, $i = j$ or $i \neq j$. If $i \neq j$, then it is either the case that $i < j$ (Lemma \ref{lemma3Q}) or $i > j$ (Lemma \ref{lemma2Q}). We can rely on the symmetrical nature of this relationship for all ballots $i,j$ and focus on proving Lemma \ref{lemma3Q}. The proof for Lemma \ref{lemma2Q} can be derived from the truth of Lemma \ref{lemma3Q}. See Appendix \ref{apendix:a} for a detailed explanation and expanded proof.

Similarly to the proof for Lemma \ref{lemma1Q}, we first define and prove Lemma \ref{lemma3Qa}. A proof of Lemma \ref{lemma3Qa} along with the contrapositive of Definition \ref{LearnImpliesPropose} (\textit{Chosen Implies Propose}), proves, using transitivity of implication, Lemma \ref{lemma3Q}.

\begin{customlemma}{4.3'}
\label{lemma3Qa}
If $v$ is chosen in ballot $i$, then no value $w \neq v$  could be proposed in any ballot $j > i$
\end{customlemma}

\begin{proof} We prove \ref{lemma3Qa} by induction on $r \geq 1$ for all ballots $j \geq i+r$.  Assume that value $v$ \textit{has been chosen in ballot $i$}, and call the Consecutive Quorum of acceptors that chose this value $Q_i$.

\noindent \textbf{Base Case:} $r = 1$.  Suppose some proposer proposed a value in ballot $j = i+1$. For a value to be proposed in ballot $j=i+1$, a quorum of acceptors must have sent corresponding $promise$ messages to the proposer for ballot $j$; call this quorum $Q_j$.


From the definition of a Consecutive Quorum, there must be at least one acceptor in $Q_i$ that accepted $v$ in ballot $i$. By quorum intersection, at least one acceptor in $Q_i$ is also in $Q_j$. If this overlapping acceptor was the acceptor that accepted $v$ in ballot $i$, this overlapping acceptor will contain the highest numbered proposal. There are no ballots higher than $i$ that could be reported in ballot $i+1$, thus proposer $p$ will propose the same value $v$. Otherwise, the overlapping acceptor accepted $v$ in some ballot $< i$. From the construction of a Consecutive Quorum, there are no interim ballots between this reported ballot and $i+1$ that an acceptor could have accepted any value other than $v$. 
\arxiv{Make more concise -- Therefore the highest reported proposal among the promise messages received by the proposer in ballot $i+1$ must contain the value $v$.   By the protocol, proposer $p$ will propose $v$.}
To constitute a valid Consecutive Quorum, all $accept$ messages in the quorum must have the same value. From the construction of Paxos, at most a single unique value can be proposed per ballot. If this overlapping ballot is not the highest reported ballot observed by the proposer, the highest proposal must contain the same value $v$. By the protocol, proposer $p$ will propose $v$.
\noindent \textbf{Inductive Step:} Assume that no proposer in the span of ballots $i,\dots, i+r-1$ proposed a value $w \neq v$. We show that the proposer $p$ in ballot $i+r$ will also not propose  $w \neq v$. 

The proposer $p$ in ballot  $i+r$ must obtain a valid quorum of $promise$ messages before proposing a value. From quorum intersection, there must be at least one acceptor that sends a $promise$ message to the proposer in ballot $i+r$ that also sent an $accept$ message to form $Q_i$. The overlapping acceptor must have accepted $v$. If the overlapping acceptor contained the highest reported ballot observed by the proposer, the proposer will propose $v$. From the construction of Consecutive Quorums, all values proposed in the span of ballots from $Q_i$ must be $v$. This in combination with the inductive hypothesis, implies that the only value that was proposed and could have been accepted in the span from the overlapping acceptor's ballot to $i+r-1$ is $v$. From the construction of Paxos, at most one value can be proposed per ballot. If this overlapping ballot is not the highest reported ballot observed by the proposer, the highest proposal must contain the same value $v$. By the protocol, proposer $p$ will propose $v$.

\end{proof}

We have proved Lemmas \ref{lemma1Q}, \ref{lemma2Q}, and \ref{lemma3Q}, and thus prove the safety of Paxos with Consecutive Quorums.

\subsection{Impact of Consecutive Quorums on Liveness}

First, Consecutive Quorums does not change the behavior of acceptors and proposers, nor does it change how learners receive messages from other agents. This implies that the set of all possible executions of Paxos is the same with and without Consecutive Quorums. Second, Consecutive Quorums only weakens the criterion to choose and learn a value. Thus, given any execution of Classic Paxos that eventually learns a value, a value is guaranteed to be eventually learned in that same execution with Consecutive Quorums. This proves that Consecutive Quorums do not impact the liveness of Paxos.

\subsection{Faster Learning With Consecutive Quorums}
Consecutive Quorums allows for values to possibly be learned in a faster and more flexible manner than traditional majority quorums. In an ideal case, a proposer can successfully propose a value that will be accepted by a majority of acceptors in the course of a single ballot. If the network is faulty, or replicas are unstable, execution may result in rapid view changes or expensive communication. The weakened version of learning that Consecutive Quorums lends to Paxos greater flexibility.

\section{Additional uses of consecutive ballots} \label{DiscussionSection}


We have introduced and proven one way in which consecutive ballots can be used to weaken learning in Paxos. Consecutive ballots can also be used to weaken the criterion for proposing values. A proposer can immediately propose a value after receiving a $promise$ message from the consecutive previous ballot. In fact, Heidi Howard in her thesis \cite{ConsensusRevised} observed that, once a proposer in ballot $n$ receives a promise message from some ballot $m<n$, that any promise messages from ballots $<m$ contain no new information. In the case that ballots are discrete, this can be extended such that if $m$ and $n$ are consecutive, the proposer can safely proceed to Phase 2 immediately. This can take place even before the proposer has a full quorum of promise messages, which is the original criterion in Paxos. 

The Classic Paxos criterion to propose a value is the following:
\begin{definition}
\label{PaxosPropose}[Normal Proposal]

\begin{align*}
    \exists Q_i .\; &[(\forall \alpha \in Q_i .\; \promise{i}{\bot}{\alpha} \in msgs_\pi) \\
    &\lor \; ((\exists \alpha \in Q_i .\; \promise{i}{\proposal{j}{v}}{\alpha} \in msgs_\pi)  \\
    &\quad\land \; (\forall \gamma \in Q_i .\; \promise{i}{\proposal{j'}{w}}{\gamma} \in msgs_\pi \land j \geq j')) ]  \implies  \propose{i}{v}{\pi}
\end{align*}
\end{definition}

A proposer $\pi$ in Classic Paxos can proceed to Phase 2 when the Normal criterion for proposing a value is met \textbf{or} the following:

\begin{definition}
\label{consecP}[Consecutive Proposal]

\[  \exists \alpha .\; \promise{i}{\proposal{i-1}{v}}{\alpha} \in msgs_\pi \implies \propose{i}{v}{\pi} \]
\end{definition}

In addition to what Howard proposed, we show how to extend the definition of Consecutive Proposals. As observed in Fast Paxos\cite{FastPaxos} for reconfiguration, receiving an $accept$ message relays the information that a $promise$ message from the same acceptor would also contain. This can be used to extend the definition of a Consecutive Proposal. A $promise$ message contains the proposal that the source acceptor has most recently accepted a value in. With our model, an agent can operate as multiple roles and could receive both $promise$ and $accept$ messages. This model of an agent operating as multiple roles is standard in practical settings \cite{Ironfleet,PaxosLive,libpaxos,Spanner}. Otherwise, an acceptor could broadcast the $accept$ message to all agents rather than just to the learners. An $accept$ message sent to agent $\theta$ from an acceptor for value $v$ in ballot $i-1$ will contain information identical to that sent by that same acceptor to agent $\theta$ in a $promise$ message in ballot $i$. We can use this to further extend the definition of a Consecutive Proposal to be valid if the proposer observes a $promise$ message from an acceptor that accepted a value in the previous ballot or an $accept$ message from an acceptor in the previous ballot. For proposer $\pi$, this is formally defined as:
\begin{definition}
\label{consecP_wAccept}[Consecutive Proposal Extended] 
    
\[ \exists \alpha .\; \promise{i}{\proposal{i-1}{v}}{\alpha} \in msgs_\pi 
\lor \accept{i-1}{v}{\alpha} \in msgs_\pi \implies \propose{i}{v}{\pi}  \]

\end{definition}

To provide a brief intuition concerning the safety of Consecutive Proposals, consider an execution of Paxos using Consecutive Proposals where ballot number $k > 0$ is the first instance of a Consecutive Proposal. The proposal in ballot $k-1$ must have been a Normal Proposal. The Normal Proposal in ballot $k-1$ was made based on a majority quorum of $promise$ messages. The first $k-1$ ballots are no different than Classic Paxos, so safety is preserved in this span. From the definition of Consecutive Proposals, the proposer in ballot $k$ must propose the same value that was proposed in the Normal Proposal from ballot $k-1$. Following ballot $k$ there could be a series of $i$ additional ballots that have Consecutive Proposals. This series of $i$ Consecutive Proposals do not have a full quorum of $promise$ messages. From the construction of a Consecutive Proposal, each Consecutive Proposal in the span of $k$ to $k+i$ can only propose the same value as the previous ballot. This value is the same as the value proposed in the most recent Normal Proposal from ballot $k-1$. Even though these proposals do not have a majority of acceptors promising not to participate in a previous ballot, there was a majority of acceptors  that did promise in ballot $k-1$. Even if there is a late arriving message for a different value, it must have come from a proposal prior to  ballot $k-1$. At least a majority of acceptors will ignore it based on the promise in ballot $k-1$. Since a majority of acceptors will ignore any previous messages, a different value can not be chosen in this span. There will never exist a majority of acceptors that accept a different value in this interim. The first ballot after the span of Consecutive Proposals is a Normal Proposal, ballot $k+i+1$. The proposer in ballot $k+i+1$ will obtain a majority quorum of $promise$ messages, and choose to propose the value associated with the highest reported ballot. If a value has already been chosen, from quorum intersection there will be at least one acceptor in common. If the value was chosen before ballot $k-1$, then all proposals in between must contain the same value. If the value was learned on or after ballot $k-1$, the series of Consecutive Proposals only propose this same value. This covers the case that the reported proposal from this overlapping acceptor is not the largest, no different value could have been proposed in a higher ballot. \arxiv{Clarify Reviewer 2 -- emphasise a value could 'still' be chosen but only the same value will be proposed so safe} If no value has previously been chosen then no matter what value is proposed, there now exists a full quorum of acceptors who promise to ignore any messages from previous ballots, preventing a different previous value being chosen. This prevents a potentially different value being learned as the result of some belated Consecutive Proposal. After this pattern of Normal-Consecutive-Normal is complete, there again exists a majority of acceptors who promise to not participate in a previous ballot less than $k+i+1$.


\section{Formal Verification}\label{formalVerification}
\subsection{Model}
Consecutive Quorums and Consecutive Proposals are not mutually exclusive. We define \textit{Consecutive Paxos} as a protocol which is a combination of Consecutive Quorums and Consecutive Proposals.

To validate our confidence in Consecutive Paxos, we make use of formal verification to produce a mechanically-checked proof. Not only do machine-checked proofs allow for greater confidence of correctness, but they often help in the construction of more refined paper proofs. Model checkers like TLA+ \cite{TLA} are standard tools used to show the correctness of distributed protocols. TLA+ certainly provides stronger guarantees than  paper proofs, but still requires the protocol specification to be trusted.


Formal verification helps in the feasibility of proving more complex systems. In this work, we build on top of the existing work from the IronFleet project \cite{Ironfleet}. The IronFleet project showed that implementations of complex distributed systems can be formally verified. IronFleet uses the Dafny programming language \cite{Dafny}, which in turn uses the Z3 SMT solver \cite{Z3} to automate the verification process. IronFleet relies on refinement proofs between various levels of abstraction; e.g., between an actual implementation and an abstract high-level specification. The benefit of using IronFleet, is that the protocol specification need not be trusted; the only trust is in the high level specification. 

We modify IronRSL, included in IronFleet, a proven replicated state machine library based on an optimized version of Multi-Paxos, that supports batching, dynamic view-changes, log truncation, and much more. We have modified the protocol specification to include both Consecutive Quorums and Consecutive Proposals. In keeping with the methodology of IronFleet, we prove a refinement between the adjusted Paxos protocol to the high-level specification of the replicated state machine library.

Using the techniques from Ironfleet, we prove that the replicated state machine built on the modified Multi-Paxos maintains linearizability of client requests. Any correct SMR protocol must satisfy this high-level specification. The refinement proof establishes a refinement between a valid sequence of Paxos protocol states and a corresponding sequence of high-level system specification states of the SMR library.





\subsection{Formal Verification of Consecutive Quorums and Consecutive Proposals}
Even though both Consecutive Quorums and Consecutive Proposals were discussed separately for the sake of brevity, they are not mutually exclusive. Paxos can be weakened in both ways simultaneously, changing the criteria to learn as well as the criteria to propose values, while retaining the safety property. In fact, both in TLA+ and with Formal Verification we showed the correctness of Paxos when modifying the protocol with both Consecutive Quorums and Consecutive Proposals. The TLA+ proof can be found in Appendix \ref{apendix:consecP+Q}.

We modified the existing refinement proof to show a refinement from the updated protocol specification to the high-level system specification of the SMR library. In the IronRSL model, replicas act as all three agents, proposers, acceptors and learners, and broadcast all messages. This allows for proposers to locally receive $accept$ messages needed for fully leveraging Consecutive Proposals. The crux of this refinement proof is identical to Lemma \ref{lemma3Qa}. We needed to show that once a valid quorum of acceptors met the updated criteria to choose a value, that any future proposal must have a matching value. The proof was a combination of the inductive proofs for Lemmas \ref{lemma3Qa} and the proof for Consecutive Proposals. Additionally, we needed to prove that a value was indeed chosen when the updated criteria for choosing a value was met, as defined for Consecutive Quorums.

\section{Conclusion}\label{conclusion}

In this paper, we rethink the fundamental building blocks that lend Paxos its correctness, by taking a closer look at the role of consecutive ballots. We show that what has long been considered an essential requirement for learning values is actually stronger than necessary. We instead propose a weaker alternative, Consecutive Quorums, that leverages the properties of consecutive ballots.

We prove that our weaker learning criterion provides the same safety and liveness guarantees as the original Paxos algorithm. In addition to our paper proof, we also provide a formally verified, mechanically-checked proof that our weaker version of Paxos indeed provides the same correctness guarantees as the original.

\bibliographystyle{ACM-Reference-Format}
\bibliography{refs}

\section*{Appendix}
  
\appendix
\section{Safety Proof Extension}\label{apendix:a}

\noindent To prove both Lemma \ref{lemma2Q} and Lemma \ref{lemma3Q} we can define and prove the following Lemmas: 

\begin{customlemma}{4.2'}
\label{lemma2Qa}
If an acceptor has accepted $v$ in ballot $i$ then no other value $w \neq v$  could be chosen in any ballot $j < i$
\end{customlemma}
By using Definition \ref{LearnImpliesAccept} (\textit{Chosen Implies Accept}) and showing that Lemma \ref{lemma2Qa} is true, with the transitivity of implication, Lemma \ref{lemma2Q} also holds. 

\begin{customlemma}{4.3'}
\label{lemma3Qa}
If $v$ is chosen in ballot $i$, then no value $w \neq v$  could be proposed in any ballot $j > i$
\end{customlemma}

Lemma \ref{lemma3Qa} is structured slightly different than the previous Lemmas (\ref{lemma1Qa} and \ref{lemma2Qa}).  A proof of Lemma \ref{lemma3Qa} along with the contrapositive of Definition \ref{LearnImpliesPropose} (\textit{Chosen Implies Propose}), proves, using transitivity of implication, Lemma \ref{lemma3Q}.

On inspection, it appears that Lemma \ref{lemma2Q} and Lemma \ref{lemma3Q} are symmetrical. In order to take advantage of this, we show that Lemma \ref{lemma3Qa} $\implies$ Lemma \ref{lemma2Qa}. This is done by taking the contrapositive of Lemma \ref{lemma2Qa}, and using Definition \ref{AcceptImpliesPropose} (\textit{Accept Implies Propose}).


These Lemmas deal with all ballots $i,j \in \mathbb{Z^+}$, and a value $v$. 
 \begin{customlemma}{4.2'}
\label{lemma2Qa}
If an acceptor has accepted $v$ in ballot $i$ then no other value $w \neq v$ could be chosen in any ballot $j < i$
\end{customlemma}

\begin{equation}
    \exists \theta \in Acceptors .\; \accept{v}{i}{\theta} \in sent_\theta \implies (\forall w,j .\; w \neq v \land j < i \;\implies \; \neg Chosen(w,j) )
\end{equation}

\begin{customlemma}{4.3'}
\label{lemma3Qa}
If $v$ is chosen in ballot $i$, then no value $w \neq v$  could be proposed in any ballot $j > i$
\end{customlemma}

\begin{equation}
    Chosen(v,i) \implies (\forall w,j .\; w \neq v \land j > i \implies (\forall \theta \in Proposers .\; \propose{w}{j}{\theta} \notin sent_\theta) )
\end{equation}

Taking the contrapositive of Lemma \ref{lemma2Qa} results in:

\begin{customlemma}{4.2'\_Contrapositive}
\label{lemma3aContraP}
If $w \neq v$ is chosen in some ballot $j < i$, then no acceptor has accepted $v$ in ballot $i$
\end{customlemma}
\begin{equation}
    \exists w,j .\; w \neq v \land j < i \land Chosen(w,j) \implies (\forall \theta \in Acceptors .\; \accept{v}{i}{\theta} \notin sent_\theta)
\end{equation}

\noindent Applying transitivity with the contrapositive of Definition \ref{AcceptImpliesPropose} (\textit{Accept implies Propose}) to Lemma \ref{lemma3Qa} results in the following: 

\begin{customlemma}{4.3'\_Accept}
\label{lemma3QaPropos}
If $w$ is chosen in ballot $j$, then no value $v \neq w$  could be accepted in any ballot $i > j$
\end{customlemma}

\begin{equation}
    Chosen(w,j) \implies (\forall v,i .\; w \neq v \land i > j \implies (\forall \theta \in Acceptors .\; \accept{v}{i}{\theta} \notin sent_\theta) )
\end{equation}

After performing these steps, we can observe that Lemma \ref{lemma3QaPropos} is equivalent to Lemma \ref{lemma3aContraP}. This follows that by proving Lemma \ref{lemma3Qa}, we can also directly show that Lemma \ref{lemma2Qa} holds. By proving these two Lemmas, both Lemma \ref{lemma3Q} and Lemma \ref{lemma2Q} can be shown directly. Additionally, to prove both Lemma \ref{lemma2Q} and Lemma \ref{lemma3Q}, it is sufficient to just prove Lemma \ref{lemma3Qa}.

Taking a step back from the notations of $i,j,v,w$, Lemma \ref{lemma3QaPropos} shows that if a value is chosen in \textbf{any} ballot then no other value could be accepted in any higher ballot. Even though the notation is seems slightly different, Lemma \ref{lemma3aContraP} states that if a value was chosen in some  \textbf{single} ballot, then no other value could be accepted in any higher ballot. This shows the symetrical relationship between Lemma \ref{lemma2Q} and Lemma \ref{lemma3Q}.

\section{Consecutive Quorums And Consecutive Proposals TLA+}\label{apendix:consecP+Q}
The TLA+ Model for the combined proof of Consecutive Quroums and Consecutive Proposals is based on the TLA+ models for Flexible Paxos \footnote{https://github.com/tlaplus/Examples/tree/master/specifications/fpaxos}. The original model was modified to include classic majority quorum definitions rather than flexible. Additionally, the  model was changed for the corresponding Consecutive Quorums and Consecutive Proposals definitions.

\tlatex
\setboolean{shading}{true}
\@x{}\moduleLeftDash\@xx{ {\MODULE} ConsecP\_AND\_Q}\moduleRightDash\@xx{}%
\@x{ {\EXTENDS} Integers ,\, Naturals ,\, TLC}%
\@pvspace{8.0pt}%
\@x{ {\CONSTANT} Value ,\, Acceptor ,\, Quorum}%
\@pvspace{8.0pt}%
 \@x{ {\ASSUME} QuorumAssumption \.{\defeq} \.{\land} \A\, Q \.{\in} Quorum
 \.{:} Q \.{\subseteq} Acceptor}%
 \@x{\@s{145.19} \.{\land} \A\, Q1 ,\, Q2 \.{\in} Quorum \.{:} Q1 \.{\cap} Q2
 \.{\neq} \{ \}}%
\@pvspace{8.0pt}%
\@x{ Ballot \.{\defeq}\@s{4.1} Int}%
\@pvspace{8.0pt}%
\@x{ None \.{\defeq} {\CHOOSE} v \.{:} v \.{\notin} Ballot}%
\@pvspace{16.0pt}%
 \@x{ Message \.{\defeq}\@s{20.5} [ type\@s{2.83} \.{:} \{\@w{1a} \} ,\,
 bal\@s{1.93} \.{:} Ballot ]}%
 \@x{\@s{49.10} \.{\cup}\@s{15.97} [ type\@s{2.83} \.{:} \{\@w{1b} \} ,\, acc
 \.{:} Acceptor ,\, bal \.{:} Ballot ,\,}%
 \@x{\@s{78.97} mbal \.{:} Ballot \.{\cup} \{ \.{-} 1 \} ,\, mval \.{:} Value
 \.{\cup} \{ None \} ]}%
 \@x{\@s{49.10} \.{\cup}\@s{15.97} [ type\@s{2.83} \.{:} \{\@w{2a} \} ,\,
 bal\@s{1.93} \.{:} Ballot ,\, val \.{:} Value ]}%
 \@x{\@s{49.10} \.{\cup}\@s{15.97} [ type\@s{2.83} \.{:} \{\@w{2b} \} ,\, acc
 \.{:} Acceptor ,\, bal \.{:} Ballot ,\, val \.{:} Value ]}%
\@pvspace{8.0pt}%
\@x{ {\VARIABLE} maxBal ,\,}%
\@x{\@s{46.84} maxVBal ,\,}%
\@x{\@s{46.84} maxVal ,\,}%
\@x{\@s{46.84} msgs}%
\@pvspace{24.0pt}%
 \@x{ vars \.{\defeq} {\langle} maxBal ,\, maxVBal ,\, maxVal ,\, msgs
 {\rangle}}%
\@pvspace{8.0pt}%
 \@x{ TypeOK \.{\defeq} \.{\land} maxBal \.{\in} [ Acceptor \.{\rightarrow}
 Ballot \.{\cup} \{ \.{-} 1 \} ]}%
 \@x{\@s{56.14} \.{\land} maxVBal \.{\in} [ Acceptor \.{\rightarrow} Ballot
 \.{\cup} \{ \.{-} 1 \} ]}%
 \@x{\@s{56.14} \.{\land} maxVal \.{\in} [ Acceptor \.{\rightarrow} Value
 \.{\cup} \{ None \} ]}%
\@x{\@s{56.14} \.{\land} msgs \.{\subseteq} Message}%
\@pvspace{16.0pt}%
 \@x{ Init \.{\defeq} \.{\land} maxBal \.{=} [ a \.{\in} Acceptor \.{\mapsto}
 \.{-} 1 ]}%
 \@x{\@s{35.70} \.{\land} maxVBal \.{=} [ a \.{\in} Acceptor \.{\mapsto} \.{-}
 1 ]}%
 \@x{\@s{35.70} \.{\land} maxVal \.{=} [ a \.{\in} Acceptor \.{\mapsto} None
 ]}%
\@x{\@s{35.70} \.{\land} msgs \.{=} \{ \}}%
\@pvspace{16.0pt}%
\@x{ Send ( m ) \.{\defeq} msgs \.{'} \.{=} msgs \.{\cup} \{ m \}}%
\@pvspace{16.0pt}%
 \@x{ Phase1a ( b ) \.{\defeq} \.{\land} Send ( [ type \.{\mapsto}\@w{1a} ,\,
 bal \.{\mapsto} b ] )}%
 \@x{\@s{69.22} \.{\land} {\UNCHANGED} {\langle} maxBal ,\, maxVBal ,\, maxVal
 {\rangle}}%
\@pvspace{8.0pt}%
\@x{ Phase1b ( a ) \.{\defeq} \.{\land} \E\, m \.{\in} msgs \.{:}}%
\@x{\@s{84.43} \.{\land} m . type \.{=}\@w{1a}}%
\@x{\@s{84.43} \.{\land} m . bal \.{>} maxBal [ a ]}%
 \@x{\@s{84.43} \.{\land} maxBal \.{'} \.{=} [ maxBal {\EXCEPT} {\bang} [ a ]
 \.{=} m . bal ]}%
 \@x{\@s{84.43} \.{\land} Send ( [ type \.{\mapsto}\@w{1b} ,\, acc \.{\mapsto}
 a ,\, bal \.{\mapsto} m . bal ,\,}%
 \@x{\@s{128.29} mbal \.{\mapsto} maxVBal [ a ] ,\, mval \.{\mapsto} maxVal [
 a ] ] )}%
\@x{\@s{69.22} \.{\land} {\UNCHANGED} {\langle} maxVBal ,\, maxVal {\rangle}}%
\@pvspace{16.0pt}%
\@x{ Clause1 ( b ,\, v ) \.{\defeq} \E\, Q \.{\in} Quorum \.{:}}%
 \@x{\@s{89.63} \.{\LET} Q1b \.{\defeq} \{ m \.{\in} msgs\@s{6.20} \.{:}
 \.{\land} m . type \.{=}\@w{1b}}%
\@x{\@s{211.35} \.{\land} m . acc \.{\in} Q}%
\@x{\@s{211.35} \.{\land} m . bal\@s{1.57} \.{=} b \}}%
 \@x{\@s{114.13} Q1bv \.{\defeq} \{ m \.{\in} Q1b \.{:} m . mbal \.{\geq} 0
 \}}%
 \@x{\@s{89.63} \.{\IN}\@s{4.09} \.{\land} \A\, a \.{\in} Q \.{:} \E\, m
 \.{\in} Q1b \.{:} m . acc \.{=} a}%
\@x{\@s{114.13} \.{\land} \.{\lor} Q1bv \.{=} \{ \}}%
\@x{\@s{129.34} \.{\lor} \E\, m \.{\in} Q1bv \.{:}}%
\@x{\@s{144.55} \.{\land} m . mval \.{=} v}%
 \@x{\@s{144.55} \.{\land} \A\, mm \.{\in} Q1bv \.{:} m . mbal \.{\geq} mm .
 mbal}%
\@pvspace{16.0pt}%
 \@x{ Clause2 ( b ,\, v ) \.{\defeq}\@s{4.1} \E\, m \.{\in} msgs \.{:}
 \.{\lor} \.{\land} m . type\@s{3.34} \.{=}\@w{1b}}%
\@x{\@s{153.13} \.{\land} m . mbal\@s{0.51} \.{=} b \.{-} 1}%
\@x{\@s{153.13} \.{\land} m . mval \.{=} v}%
\@x{\@s{153.13} \.{\land} m . bal \.{=} b}%
\@x{\@s{142.02} \.{\lor} \.{\land} m . type \.{=}\@w{2b}}%
\@x{\@s{153.13} \.{\land} m . bal\@s{0.51} \.{=} b \.{-} 1}%
\@x{\@s{153.13} \.{\land} m . val \.{=} v}%
\@pvspace{24.0pt}%
\@x{ NormalProposal ( b ,\, v ) \.{\defeq} Clause1 ( b ,\, v )}%
\@pvspace{8.0pt}%
 \@x{ ConsecProposal ( b ,\, v )\@s{1.93} \.{\defeq} {\lnot} Clause1 ( b ,\, v
 ) \.{\land}\@s{4.1} Clause2 ( b ,\, v )}%
\@pvspace{8.0pt}%
\@x{ Phase2a ( b ,\, v ) \.{\defeq}}%
 \@x{\@s{8.2} \.{\land} {\lnot} \E\, m \.{\in} msgs \.{:} m . type
 \.{=}\@w{2a} \.{\land} m . bal \.{=} b}%
 \@x{\@s{8.2} \.{\land} ( NormalProposal ( b ,\, v ) \.{\lor} ConsecProposal (
 b ,\, v ) )}%
 \@x{\@s{8.2} \.{\land} Send ( [ type \.{\mapsto}\@w{2a} ,\, bal \.{\mapsto} b
 ,\, val \.{\mapsto} v ] )}%
 \@x{\@s{8.2} \.{\land} {\UNCHANGED} {\langle} maxBal ,\, maxVBal ,\, maxVal
 {\rangle}}%
\@pvspace{24.0pt}%
 \@x{ Phase2b ( a ) \.{\defeq} \E\, m \.{\in} msgs \.{:} \.{\land} m . type
 \.{=}\@w{2a}}%
\@x{\@s{128.83} \.{\land} m . bal \.{\geq} maxBal [ a ]}%
 \@x{\@s{128.83} \.{\land} maxBal \.{'} \.{=} [ maxBal {\EXCEPT} {\bang} [ a ]
 \.{=} m . bal ]}%
 \@x{\@s{128.83} \.{\land} maxVBal \.{'} \.{=} [ maxVBal {\EXCEPT} {\bang} [ a
 ] \.{=} m . bal ]}%
 \@x{\@s{128.83} \.{\land} maxVal \.{'} \.{=} [ maxVal {\EXCEPT} {\bang} [ a ]
 \.{=} m . val ]}%
 \@x{\@s{128.83} \.{\land} Send ( [ type \.{\mapsto}\@w{2b} ,\, acc
 \.{\mapsto} a ,\,}%
 \@x{\@s{168.59} bal \.{\mapsto} m . bal ,\, val\@s{2.06} \.{\mapsto} m . val
 ] )}%
\@pvspace{8.0pt}%
 \@x{ Next \.{\defeq} \.{\lor} \E\, b\@s{0.64} \.{\in} Ballot \.{:} \.{\lor}
 Phase1a ( b )}%
\@x{\@s{112.86} \.{\lor} \E\, v \.{\in} Value \.{:} Phase2a ( b ,\, v )}%
 \@x{\@s{39.83} \.{\lor} \E\, a \.{\in} Acceptor \.{:} Phase1b ( a ) \.{\lor}
 Phase2b ( a )}%
\@pvspace{8.0pt}%
\@x{ Spec\@s{1.46} \.{\defeq} Init \.{\land} {\Box} [ Next ]_{ vars}}%
\@pvspace{8.0pt}%
\@x{ NotIn ( Q ,\, a ) \.{\defeq} a \.{\notin} Q}%
\@pvspace{8.0pt}%
 \@x{ ConsecQ ( Q ,\, b ,\, v ) \.{\defeq}\@s{4.1} \A\, r \.{\in} Acceptor
 \.{:} \.{\land} NotIn ( Q ,\, r )}%
\@x{\@s{174.49} \.{\land} \E\, m \.{\in} msgs \.{:}}%
\@x{\@s{192.82} \.{\land} m . type \.{=}\@w{2b}}%
\@x{\@s{192.82} \.{\land} m . acc \.{=} r}%
 \@x{\@s{192.82} \.{\implies} \.{\lor} ( \A\, q \.{\in} Q \.{:} \E\, n \.{\in}
 msgs \.{:} \.{\land} n . type \.{=}\@w{2b}}%
\@x{\@s{323.41} \.{\land} n . acc\@s{5.42} \.{=} q}%
\@x{\@s{323.41} \.{\implies} m . bal \.{<} n . bal )}%
 \@x{\@s{208.38} \.{\lor} ( \A\, q \.{\in} Q \.{:} \E\, n \.{\in} msgs \.{:}
 \.{\land} n . type \.{=}\@w{2b}}%
\@x{\@s{323.41} \.{\land} n . acc\@s{2.86} \.{=} q}%
\@x{\@s{323.41} \.{\implies} n . bal \.{<} m . bal )}%
\@pvspace{24.0pt}%
 \@x{ Sent2b ( a ,\, v ,\, b ) \.{\defeq} \E\, m \.{\in} msgs \.{:} \.{\land}
 m . type \.{=}\@w{2b}}%
\@x{\@s{145.63} \.{\land} m . acc \.{=} a}%
\@x{\@s{145.63} \.{\land} m . val\@s{1.06} \.{=} v}%
\@x{\@s{145.63} \.{\land} m . bal\@s{1.57} \.{=} b}%
\@pvspace{16.0pt}%
 \@x{ Sent2bNoBal ( a ,\, v ) \.{\defeq} \E\, m \.{\in} msgs \.{:} \.{\land} m
 . type \.{=}\@w{2b}}%
\@x{\@s{161.94} \.{\land} m . acc \.{=} a}%
\@x{\@s{161.94} \.{\land} m . val\@s{1.06} \.{=} v}%
\@pvspace{8.0pt}%
 \@x{ Sent2a ( v ,\, b ) \.{\defeq} \E\, m \.{\in} msgs \.{:} \.{\land} m .
 type \.{=}\@w{2a}}%
\@x{\@s{134.28} \.{\land} m . val \.{=} v}%
\@x{\@s{134.28} \.{\land} m . bal\@s{0.51} \.{=} b}%
\@pvspace{16.0pt}%
 \@x{ NormalBeforeSpec ( b ,\, v ) \.{\defeq} \.{\land} ConsecProposal ( b ,\,
 v )}%
\@x{\@s{123.88} \.{\land} \E\, x \.{\in} Ballot \.{:} \.{\land} x \.{<} b}%
\@x{\@s{196.88} \.{\land} NormalProposal ( x ,\, v )}%
\@pvspace{8.0pt}%
 \@x{ SpecSameValAsLastNorm ( b ,\, v ) \.{\defeq} \.{\land} ConsecProposal (
 b ,\, v )}%
\@x{\@s{155.95} \.{\land} \E\, x \.{\in} Ballot \.{:}}%
\@x{\@s{188.32} \.{\land} x \.{<} b}%
\@x{\@s{188.32} \.{\land} NormalProposal ( x ,\, v )}%
 \@x{\@s{188.32} \.{\land} \A\, y \.{\in} Ballot \.{:} ( y \.{>} x \.{\land} y
 \.{\leq} b ) \.{\implies} ConsecProposal ( y ,\, v )}%
\@pvspace{16.0pt}%
 \@x{ AgreedConsecQ ( v ,\, b ) \.{\defeq} \E\, Q \.{\in} Quorum \.{:}\@s{4.1}
 \.{\land} \A\, a \.{\in} Q \.{:} Sent2bNoBal ( a ,\, v )}%
\@x{\@s{189.74} \.{\land} \E\, a \.{\in} Q \.{:} Sent2b ( a ,\, v ,\, b )}%
\@x{\@s{189.74} \.{\land} ConsecQ ( Q ,\, b ,\, v )}%
\@pvspace{16.0pt}%
 \@x{ Agreed ( v ,\, b ) \.{\defeq} \E\, Q \.{\in} Quorum \.{:} \A\, a \.{\in}
 Q \.{:} Sent2b ( a ,\, v ,\, b )}%
 \@x{ NoFutureProposal ( v ,\, b ) \.{\defeq} \A\, v2 \.{\in} Value \.{:} \A\,
 b2 \.{\in} Ballot \.{:} ( b2 \.{>} b \.{\land} Sent2a ( v2 ,\, b2 ) )
 \.{\implies} v \.{=} v2}%
\@pvspace{8.0pt}%
 \@x{ SafeValue \.{\defeq} \A\, v \.{\in} Value \.{:} \A\, b \.{\in} Ballot
 \.{:} AgreedConsecQ ( v ,\, b )\@s{4.1} \.{\implies} NoFutureProposal ( v
 ,\, b )}%
\@pvspace{8.0pt}%
 \@x{ SpecSafety \.{\defeq} \A\, v \.{\in} Value \.{:} \A\, b \.{\in} Ballot
 \.{:} ConsecProposal ( b ,\, v )}%
 \@x{\@s{65.26} \.{\implies} NormalBeforeSpec ( b ,\, v ) \.{\land}
 SpecSameValAsLastNorm ( b ,\, v )}%
\@pvspace{8.0pt}%
\@x{}\bottombar\@xx{}%
\setboolean{shading}{false}

\end{document}